\documentclass[sigconf,authorversion]{acmart} 

\def\BibTeX{{\rm B\kern-.05em{\sc i\kern-.025em b}\kern-.08emT\kern-.1667em\lower.7ex\hbox{E}\kern-.125emX}}
    
\newcommand{\vc}[1]{\boldsymbol{#1}}
\newcommand{\Co}{\mathcal{C}}
\newcommand{\expn}{\mathbb{E}}

\begin{document}

\title{Nearest neighbor decoding for Tardos fingerprinting codes}

\author{Thijs Laarhoven}
\email{mail@thijs.com}
\orcid{0000-0002-2369-9067}
\affiliation{%
  \institution{Eindhoven University of Technology}
  \city{Eindhoven}
  \country{The Netherlands}
}

\begin{abstract}
Over the past decade, various improvements have been made to Tardos' collusion-resistant fingerprinting scheme [Tardos, STOC 2003], ultimately resulting in a good understanding of what is the minimum code length required to achieve collusion-resistance. In contrast, decreasing the cost of the actual decoding algorithm for identifying the potential colluders has received less attention, even though previous results have shown that using joint decoding strategies, deemed too expensive for decoding, may lead to better code lengths. Moreover, in dynamic settings a fast decoder may be required to provide answers in real-time, further raising the question whether the decoding costs of score-based fingerprinting schemes can be decreased with a smarter decoding algorithm.

In this paper we show how to model the decoding step of score-based fingerprinting as a nearest neighbor search problem, and how this relation allows us to apply techniques from the field of (approximate) nearest neighbor searching to obtain decoding times which are sublinear in the total number of users. As this does not affect the encoding and embedding steps, this decoding mechanism can easily be deployed within existing fingerprinting schemes, and this may bring a truly efficient joint decoder closer to reality.

Besides the application to fingerprinting, similar techniques can be used to decrease the decoding costs of group testing methods, which may be of independent interest.
\end{abstract}




\maketitle

\newpage
\section{Introduction}
\label{sec:intro}

\subsection{Digital fingerprinting}
Fingerprinting techniques for digital content provide a way for copyright holders to uniquely mark each copy of their content, to prevent unauthorized redistribution of this content: if a digital ``pirate'' nevertheless decides to publicly share his (fingerprinted) content with others, the owner of the content can obtain this copy, extract the fingerprint, link it to the responsible user, and take appropriate steps. Digital pirates may try to prevent being caught by collaborating, and forming a mixed copy of the content from their individual copies, thus mixing up the embedded fingerprint as well. To guarantee that collusions of pirates cannot get away with this, \textit{collusion-resistant fingerprinting schemes} are needed.

Mathematically speaking, collusion-resistant fingerprinting can be modeled as follows. First, the content owner generates code words $\vc{x}_j \in \{0,1\}^{\ell}$ for $j = 1, \dots, n$, corresponding to fingerprints for the $n$ users, where each of the $\ell$ columns defines one segment of the content. Then, a collusion $\Co$ of $c$ colluders applies a mixing strategy to their code words $\{\vc{x}_j\}_{j \in \Co}$ to form a new pirate copy $\vc{y}$. Here the critical condition we impose on this mixing procedure is the \textit{marking assumption}, stating that if $x_{j,i} = b$ for all $j \in \Co$, then also $y_i = b$, for $b \in \{0,1\}$. Finally, the owner of the content obtains $\vc{y}$, applies a decoding algorithm to $\vc{y}$ and all the code words $\{\vc{x}_j\}_{j=1}^n$, and outputs a subset of user indices $j$. This method is successful if, with high probability over the randomness in the code generation and mixing strategy, the decoding algorithm outputs (a subset of) the colluders, without incriminating any innocent, legitimate users.

\subsection{Related work}
In the late 1990s, Boneh--Shaw~\cite{boneh98} were the first to design a somewhat practical, combinatorial solution for collusion-resistant fingerprinting. Their construction based on error-correcting codes achieved high success probabilities with a code length of $\ell = O(c^4 \log n)$, i.e.\ scaling logarithmically in the often large number of users $n$, and quarticly in the number of colluders $c$. The milestone work of Tardos~\cite{tardos03} later improved upon this with a code length $\ell = O(c^2 \log n)$, and he proved that this quadratic scaling in $c$ is optimal. Later work focused on bringing down the leading constants in Tardos' scheme~\cite{skoric08b, blayer08, furon08, furon09c, meerwald12, oosterwijk13, furon14, skoric15}, leading to an optimal code length of $\ell \propto \frac{1}{2} \pi^2 c^2 \ln n$ for the original (symmetric) Tardos score function~\cite{skoric08, laarhoven14dcc, laarhoven13ihmmsec}, and an optimal overall code length of $\ell \propto 2 c^2 \ln n$ when using further improved decoders~\cite{oosterwijk13, oosterwijk13b, desoubeaux13, laarhoven14ihmmsec, furon14}. 

The main focus of most literature on collusion-resistant fingerprinting has been on decreasing $\ell$ -- the shorter the fingerprints, the faster the colluders can be traced. Other aspects of these fingerprinting schemes, however, have received considerably less attention in the literature, and in particular the decoding procedure is often neglected. Indeed, with $n$ users and a code length of $\ell = O(c^2 \log n)$, the decoding time is commonly $O(\ell \cdot n) = O(c^2 n \log n)$ for linear decoders, and up to $O(c^2 n^k \log n)$ for \textit{joint decoders}, attempting to decode to groups of $k \leq c$ colluders simultaneously. In particular, past work has shown that joint decoders achieve superior performance to simple decoders~\cite{amiri09, charpentier09, furon09b, meerwald12, furon12, berchtold12, laarhoven14ihmmsec, laarhoven16phd}, but are often considered infeasible due to their high decoding complexity. Moreover, in \textit{dynamic} settings~\cite{laarhoven13tit, laarhoven15ihmmsec} where decisions about the accusation of users need to be made swiftly, an efficient decoding method is even more critical. Techniques that can speed up the decoding procedure may therefore be useful for further improving these schemes in practice.

\subsection{Contributions}
In this work, we study how the decoding method in the score-based fingerprinting framework can be improved, leading to faster decoding times. In particular, we show that we can typically bring down the decoding costs of simple decoders from $O(\ell n)$ to $O(\ell n^{\rho})$ where $\rho \leq 1$ is determined by $c$ and the instantiation of the score-based framework. This usually comes at a higher space requirement for indexing the code words in a query-efficient data structure, although arbitrary trade-offs between the space and query time can be obtained by tweaking the parameters.

To obtain these improved results, we show that we can model the decoding procedure of Tardos-like fingerprinting as a high-dimensional \textit{nearest neighbor search} problem. This field of research studies methods of storing data in more refined data structures, such that highly similar vectors to any given query point can be found faster than with a linear search through the data. Applying practical, state-of-the-art techniques from this area, such as the locality-sensitive hashing mechanisms of~\cite{charikar02, andoni15cp} and the more recent locality-sensitive filtering of \cite{becker16lsf, andoni17}, this allows us to obtain faster, sublinear decoding times, both in theory and in practice. We give a recipe how to apply these techniques to any score-based fingerprinting framework, and give an explicit, detailed analysis of what happens when we use these techniques in combination with the symmetric score function of \v{S}kori\'{c}--Katzenbeisser--Celik~\cite{skoric08}. 

To illustrate these results for the symmetric score function, let us provide some explicit decoding costs for the case of $c = 2$. We obtain a decoding time of $O(\ell n^{3/4})$ without using additional memory; a decoding cost of $O(\ell n^{1/3})$ when using $O(\ell n^{4/3})$ memory; and a subpolynomial decoding cost $\ell n^{o(1)}$ when using up to $O(\ell n^4)$ storage. For larger $c$ the improvement becomes smaller, and in the limit of large $c$ nearest neighbor techniques offer only minor improvements over existing decoding schemes -- the main benefit lies in defending against small or moderate collusion sizes.

\subsection{Outline.} 
The remainder of this paper is organized as follows. In Section~\ref{sec:pre} we first introduce notation, we describe the score-based fingerprinting framework, and we cover basics on nearest neighbor searching. Section~\ref{sec:main} describes how to apply nearest neighbor techniques to fingerprinting, and analyzes the theoretical impact of this application. Section~\ref{sec:exp} covers basic experiments, to illustrate the potential effects of these techniques in practice, and Section~\ref{sec:disc} finally discusses other aspects of our proposed improvement.

\newpage
\section{Preliminaries}
\label{sec:pre}

\begin{table}
  \caption{Notation used throughout the paper. The first five rows indicate how concepts in fingerprinting translate to concepts in nearest neighbor searching.}
  \label{tab:not}
  \begin{tabular}{clccl}
    \toprule
    & FP terminology & & & NNS terminology \\
    \midrule
    $n$ & Number of users & $\sim$ & $n$ & Number of points \\
    $\ell$ & Code length & $\sim$ & $d$ & Dimension of data \\
    $\vc{x}_j$ & Code word & $\sim$ & $\vc{v}_j$ & Data point \\
    $\vc{y}$ & Pirate copy & $\sim$ & $\vc{q}$ & Query point \\
    $s_j$ & User score & $\sim$ & $\langle \vc{v}_j, \vc{q} \rangle$ & Dot product \\
    \midrule
    $c$ & Colluders & & $\alpha$ & Approximation factor \\
    $\vc{p}$ & Probability vector & & $\rho_q$ & Query time exponent \\
    $g$ & Score function & & $\rho_s$ & Space exponent \\
  \bottomrule
\end{tabular}
\end{table}

\subsection{Score-based fingerprinting schemes}
We first recall the score-based fingerprinting framework, introduced by Tardos~\cite{tardos03}, and describe the model considered in this paper. The fingerprinting game consists of three phases: (1) \textit{encoding}, generating the fingerprints and embedding them in the content; (2) the \textit{collusion attack}, constructing the mixed fingerprint; and (3) \textit{decoding}, mapping the mixed fingerprint to a set of accused users.

\subsubsection{Encoding.} First, the copyright holder generates code words $\vc{x}_j \in \{0,1\}^{\ell}$ for each of the users $j = 1, \dots, n$. To do this, he first generates a probability vector $\vc{p} \in [0,1]^{\ell}$, where each coordinate $p_i$ is drawn independently from some fixed probability distribution $F$. In the original Tardos scheme, and many of its variants, a truncated version of the arcsine distribution is used, which has cumulative density function $F$ given below.
\begin{align}
    F(p) := \frac{2}{\pi} \arcsin \sqrt{p}. \qquad (0 \leq p \leq 1) \label{eq:F}
\end{align}
For small $c$ and the symmetric Tardos score function, certain discrete distributions are known to be optimal~\cite{nuida09}, leading to a better performance and shorter code lengths. For large $c$, these optimal discrete distributions converge to the arcsine distribution~\eqref{eq:F}.

After generating each entry $p_i$ from the chosen bias distribution, the code words for the users are generated as follows: for each user $j$, the $i$th entry of their codeword $x_{j,i}$ is set to $1$ with probability $p_i$, and $0$ with probability $1 - p_i$. This assignment is done independently for each $i$ and $j$. These fingerprints are then embedded in the content and sent to the users. (Note that it is crucial that the bias vector $\vc{p}$ remains secret, and is not known to the colluders during the attack.)

\subsubsection{Collusion attack.} Given a collusion $\Co$ of some size $c$, the pirates employ a strategy $\theta$, mapping their code words $\{\vc{x}_j\}_{j \in \Co}$ to a mixed copy $\vc{y} \in \{0,1\}^{\ell}$. Often the attack is modeled by a vector $\vc{\theta} = (\theta_0, \dots, \theta_c)$, where $\theta_k := \Pr(y_i = 1 | \sum_{j \in \Co} x_{j,i} = k)$. By the marking assumption, we assume $\theta_0 = 0$ and $\theta_c = 1$.

\subsubsection{Decoding.} Given $\vc{y}$, the content holder attempts to deduce who were responsible for creating this pirate copy. For this, he computes scores $s_{j,i} := g(x_{j,i}, y_i, p_i)$ for some score function $g$, and then computes the total user scores as $s_j := \sum_{i=1}^n s_{j,i}$. For well-chosen parameters, the cumulative scores $s_j$ are significantly higher for colluders than for innocent users. The actual decision whom to accuse is then made by e.g.\ setting a threshold $z$ and accusing users $j$ with $s_j > z$, or by accusing the user with the highest cumulative score. As an example, the symmetric score function of \v{S}kori\'{c}--Katzenbeisser--Celik~\cite{skoric08} is given below.
\begin{align}
    g(x, y, p) := \begin{cases} 
        +\sqrt{p/(1-p)}, & \text{if } x=0 \text{ and } y=0; \\ 
        -\sqrt{(1-p)/p}, & \text{if } x=1 \text{ and } y=0; \\ 
        -\sqrt{p/(1-p)}, & \text{if } x=0 \text{ and } y=1; \\ 
        +\sqrt{(1-p)/p}, & \text{if } x=1 \text{ and } y=1. 
    \end{cases}
\end{align}

\subsubsection{Equivalent decoding.} Since the decoding step remains equally valid after linear transformations (i.e.\ scaling all user scores by a common positive factor, or shifting all user scores by the same amount), the following scoring function is equivalent to the symmetric score function described above:
\begin{align}
    \hat{g}(x, y, p) := \begin{cases} 
        +1/\sqrt{p(1-p)}, & \text{if } x = y; \\ 
        -1/\sqrt{p(1-p)}, & \text{if } x \neq y. 
    \end{cases}
\end{align}
To see why, note that the contribution of a segment for the symmetric score function, in terms of how far the scores for a match and a difference are apart, is independent of $y$: 
\begin{align}
    g(1,1,p) - g(0,1,p) = g(0,0,p) - g(1,0,p) = \frac{1}{\sqrt{p(1-p)}}.
\end{align}
In other words, as long as the difference between a match and a difference in segment $i$ is proportional to $1/\sqrt{p_i(1-p_i)}$ (with a positive contribution for a match, and a negative contribution for a difference), this only constitutes a scaling/transformation of the scores. By scaling the scores by a factor $2$, and centering the scores at $0$, we obtain the score function $\hat{g}$. (Note that the threshold $z$ will have to be scaled and translated by the same amounts to guarantee equivalent error probabilities.)


\subsection{(Approximate) nearest neighbor searching}
Next, let us recall some definitions, techniques, and results from nearest neighbor searching (NNS). Given a data set $\{\vc{v}_1, \dots, \vc{v}_n\} \subset \mathbb{R}^d$, this problem asks to index these points in a data structure such that, when later given a query vector $\vc{q} \in \mathbb{R}^d$, one can quickly identify the nearest vector to $\vc{q}$ in the data set. To measure the performance of an NNS method, we consider the space complexity $S = O(n^{1 + \rho_s})$ and the query time complexity $T = O(n^{\rho_q})$ to process a query $\vc{q}$. Note that a naive linear search, without any indexing of the data, achieves $T = S = O(n)$ or $\rho_s = 0$ and $\rho_q = 1$. Ideally a good NNS method should achieve $\rho_q < 1$, perhaps with $\rho_s > 0$.

In this paper we restrict our attention to the NNS problem on the \textit{unit sphere}, under the $\ell_2$-norm: we assume that $\|\vc{v}_j\| = 1$ for all $j$, and $\|\vc{q}\| = 1$. The following lemma states that, if the entire data set has a small dot product $\langle \vc{v}_j, \vc{q} \rangle := \sum_{j=1}^d v_{j,d} q_j$ with $\vc{q}$, except for one near neighbor $\vc{v}_{j^*}$, which has a large dot product with $\vc{q}$, then finding this unique near neighbor can be done efficiently in sublinear time. The parameter $\alpha \geq 1$ below is commonly referred to as the \textit{approximation factor} (denoted $c$ in e.g.\ \cite{andoni17}) -- one obtains a sublinear time complexity for NNS only when either an approximate solution suffices, or there is a guarantee that the data set contains unique nearest neighbors which are a factor $\alpha$ closer (under the $\ell_2$-norm) than all other vectors in the data set.

\begin{lemma}[NNS complexities~\cite{andoni17}] \label{lem:nns}
Suppose that the data points $\vc{v}_i$ and query $\vc{q}$ have norm $1$, and we are given two guarantees:
\begin{itemize}
    \item For the nearest neighbor $\vc{v}_{j^*}$, we have $\langle \vc{v}_{j^*}, \vc{q} \rangle \geq d_1$;
    \item For all other vectors $\vc{v}_j \neq \vc{v}_{j^*}$, we have $\langle \vc{v}_j, \vc{q} \rangle \leq d_0$.
\end{itemize}
Let $\alpha = \sqrt{1 - d_0} / \sqrt{1 - d_1} \geq 1$, and let $\rho_q, \rho_s \geq 0$ satisfy:
\begin{align}
    \alpha^2 \sqrt{\rho_q} + (\alpha^2 - 1) \sqrt{\rho_s} \geq \sqrt{2\alpha^2 - 1}. \label{eq:tradeoff}
\end{align}
Then we can construct a data structure with $\tilde{O}(n^{1 + \rho_s})$ space and preprocessing time, allowing to answer any query $\vc{q}$ correctly (with high probability) with query time complexity $\tilde{O}(n^{\rho_q})$. 
\end{lemma}

To achieve the above complexities, at a high level the data structure looks as follows. Given the normalized data points $\vc{v}_j$, all lying on the unit sphere, we first sample many random vectors $\vc{r}_k$ on the sphere, and for each of these vectors we store which vectors are close to $\vc{r}_k$ in a bucket $B_k$. The key property we use here is (approximate) \textit{transitivity} of closeness on the sphere: if $\vc{x}$ and $\vc{y}$ are close, and $\vc{y}$ and $\vc{z}$ are close, then also $\vc{x}$ and $\vc{z}$ are more likely to be close than usual. In other words, if $\vc{v}_j$ is close to $\vc{q}$, and $\vc{v}_j$ is close to $\vc{r}_k$ (and contained in bucket $B_k$), then likely $\vc{q}$ will also be close to $\vc{r}_k$. Therefore, if we create and index many of these buckets $B_k$ and, given $\vc{q}$, we compute the dot products of $\vc{q}$ with the vectors $\vc{r}_k$ and only check those buckets $B_k$ for potential near neighbors for which $\langle \vc{q}, \vc{r}_k \rangle$ is large, then we may only check a small fraction of the entire data set for potential nearest neighbors, while still finding all near neighbors. (The actual state-of-the-art techniques from~\cite{andoni17} are slightly more sophisticated. For further details, we refer the reader to~\cite{andoni15cp, andoni17, becker16lsf}.)

Note that Lemma~\ref{lem:nns} only states the \textit{scaling} behavior of the time and space complexities, and does not state how large the real overhead is in practical scenarios. For actual applications of NNS techniques, we refer the reader to e.g.\ the benchmarks of~\cite{aumueller17}, comparing implementations of various NNS techniques for their practicality on real-world data sets, including data sets on the sphere.

\section{Nearest neighbor decoding}
\label{sec:main}

\subsection{Score-based decoding as an NNS problem}
To apply NNS techniques to score-based fingerprinting, let us first show how we can phrase the decoding step of score-based fingerprinting as an NNS problem on the sphere. First, we map the $n$ code words $\vc{x}_j \in \{0,1\}^{\ell}$ to $n$ data points $\vc{v}_j \in \{-1,1\}^{\ell}$ by the linear operation $\vc{v}_j = 2 \vc{x}_j - \vc{1}$: a $1$ in $\vc{x}_j$ is mapped to a $1$ in $\vc{v}_j$, and a $0$ in $\vc{x}_j$ to a $-1$ in $\vc{v}_j$. Next, given $\vc{y} \in \{0,1\}^{\ell}$, we map it to a query vector $\vc{q}$ as $q_i = (2 y_i - 1) / \sqrt{p_i (1 - p_i)}$: the entries of $\vc{q}$ are $\pm 1 / \sqrt{p_i (1 - p_i)}$, depending on the value of $y_i$. Note that the Euclidean norms of the data and query vectors are given by:
\begin{align}
    \|\vc{v}_j\| &= \sqrt{\ell}, \qquad %
    \|\vc{q}\| = \sqrt{\sum_{i=1}^{\ell} \frac{1}{p_i (1 - p_i)}}. \label{eq:norms}
\end{align}
To guarantee that all vectors are normalized, we will later have to scale everything down by $\|\vc{v}_j\|$ and $\|\vc{q}\|$ accordingly. Observe that with the modified symmetric score function $\hat{g}$, the user score $s_j$ can now be equivalently expressed in terms of dot products as follows:
\begin{align}
    s_j = \sum_{i=1}^{\ell} \hat{g}(x_{j,i},y_i,p_i) = \sum_{i=1}^{\ell} \frac{(2 x_{j,i} - 1)(2 y_i - 1)}{\sqrt{p_i (1-p_i)}} = \langle \vc{v}_j, \vc{q} \rangle.
\end{align}
Therefore, a user score $s_j$ is large iff the dot product between $\vc{v}_j$ and $\vc{q}$ is large, and $\vc{v}_j$ and $\vc{q}$ are near neighbors in space.


With the above translation in mind, we can now apply the aforementioned NNS techniques. To apply Lemma~\ref{lem:nns}, after normalization we need to provide two guarantees: 
\begin{itemize}
    \item A nearest neighbor $\vc{v}_{j^*}$ (i.e.\ a code word $\vc{x}_{j^*}$ of a colluder $j^* \in \Co$) must have a large dot product with the query vector $\vc{q}$ (i.e.\ must have a high score $s_j$);
    \item Other neighbors $\vc{v}_j$ (i.e.\ innocent users $j$) must have a small dot product with $\vc{q}$ (i.e.\ must have a low score $s_j$).
\end{itemize}
For this, we could derive similar proven bounds on $\Pr(s_j > z)$ for innocent and guilty users, as previously done in e.g.\ \cite{tardos03, skoric08, skoric08b, blayer08, laarhoven14dcc}, taking into account that the scores have been transformed. Instead let us give a slightly informal, high-level description of what these results may be. 

Let $H_0$ be the hypothesis that user $j$ is innocent, and $H_1$ the hypothesis that user $j$ is a colluder. Let $\mu_b = \expn_{\vc{p}, \vc{x}_j, \vc{y}}(s_j | H_b)$ for $b \in \{0,1\}$. By the central limit theorem, cumulative user scores are distributed approximately normally for large $\ell$, and if both variances $\sigma_b^2 = \expn_{\vc{p}, \vc{x}_j, \vc{y}}(s_j^2 | H_b) - \mu_b^2$ are small, we may conclude that with high probability these scores are closely concentrated around their means. Then we can estimate the parameters $d_0$ and $d_1$ for Lemma~\ref{lem:nns}, after normalization, as follows:
\begin{align}
    d_0 &= \frac{\mu_0}{\|\vc{v}_j\| \cdot \|\vc{q}\|}, \qquad %
    d_1 = \frac{\mu_1}{\|\vc{v}_j\| \cdot \|\vc{q}\|}.
\end{align}
Here the expressions for $\|\vc{v}_j\|$ and $\|\vc{q}\|$ follow from~\eqref{eq:norms}. Note however that both $\mu_0$ and $\mu_1$ are likely dependent on the collusion attack, and may not be known in advance, before the decoding stage.

\subsection{Two colluders}
Let us first investigate the simplest case of $c = 2$, i.e.\ having two colluders. Under the assumption that the colluders work symmetrically, there is no collusion strategy to consider: if they have the same symbol, they output this symbol, and if they receive both a $0$ and a $1$ they can choose either with equal probability (i.e.\ $\vc{\theta} = (0, \frac{1}{2}, 1)$). For $c = 2$ the best choice is to fix $p_i = \frac{1}{2}$ for all $i$, leading to uniformly random codes. This implies $\|\vc{q}\| = 2 \sqrt{\ell}$, $\mu_0 = 0$ and $\mu_1 = \ell$, resulting in $d_0 = 0$ and $d_1 = \frac{1}{2}$, as in Table~\ref{tab:num}. In Lemma~\ref{lem:nns} this leads to an approximation factor $\alpha = \sqrt{2}$, and a trade-off between the time and space exponents $\rho_q$ and $\rho_s$ of:
\begin{align}
    2 \sqrt{\rho_q} + \sqrt{\rho_s} \geq \sqrt{3}.
\end{align}
Without increasing the memory (i.e.\ for $\rho_s = 0$), we obtain $\rho_q \geq \frac{3}{4}$, i.e.\ an asymptotic query time complexity for decoding of $O(n^{3/4} \ell)$. Setting $\rho_q = \rho_s$ we obtain $\rho_q \geq \frac{1}{3}$, i.e.\ with $O(n^{4/3} \ell)$ memory, we can obtain a query complexity of $O(n^{1/3} \ell)$. With a large amount of memory and preprocessing time, we can further get a subpolynomial query time $n^{o(1)} \ell$ at the cost of $O(n^4 \ell)$ memory.

\begin{table}
  \caption{Numerical data for the interleaving attack, using the optimal discrete distributions of~\cite{nuida09}. The last three columns correspond to three extreme time--space trade-offs: (I) $\rho_q$ for $\rho_s = 0$; (II) $\rho_q$ for $\rho_q = \rho_s$; and (III) $\rho_s$ for $\rho_q = 0$.}
  \label{tab:num}
  \begin{tabular}{c|cccccc|ccc}
    \toprule
    $c$ & $\frac{\|\vc{q}\|}{\sqrt{\ell}}$ & $\frac{\mu_0}{\ell}$ & $\frac{\mu_1}{\ell}$ & $d_0$ & $d_1$ & $\alpha$ & I & II & III \\
    \midrule
    $1$ & $2.00$ & $0.00$ & $2.00$ & $0.00$ & $1.00$ & $\infty$ & $0.00$ & $0.00$ & $0.00$ \\
    $2$ & $2.00$ & $0.00$ & $1.00$ & $0.00$ & $0.50$ & $1.41$ & $0.75$ & $0.33$ & $5.00$ \\
    $3$ & $2.45$ & $0.82$ & $1.36$ & $0.33$ & $0.56$ & $1.22$ & $0.89$ & $0.50$ & $8.00$ \\
    $4$ & $2.45$ & $0.82$ & $1.22$ & $0.33$ & $0.50$ & $1.15$ & $0.94$ & $0.60$ & $15.0$ \\
    $5$ & $2.83$ & $1.26$ & $1.56$ & $0.45$ & $0.55$ & $1.11$ & $0.96$ & $0.68$ & $25.8$ \\
    $6$ & $2.83$ & $1.26$ & $1.51$ & $0.45$ & $0.54$ & $1.09$ & $0.97$ & $0.72$ & $37.6$ \\
    $7$ & $3.16$ & $1.57$ & $1.78$ & $0.50$ & $0.56$ & $1.07$ & $0.98$ & $0.77$ & $57.3$ \\
    $8$ & $3.16$ & $1.57$ & $1.75$ & $0.50$ & $0.56$ & $1.06$ & $0.99$ & $0.79$ & $75.2$ \\
    \midrule
    $\infty$ & $\infty$ & $\infty$ & $\infty$ & $0$ & $0$ & $1$ & $1$ & $1$ & $\infty$ \\
  \bottomrule
\end{tabular}
\end{table}

\subsection{More colluders}
For $c \geq 3$, the collusion strategy affects $d_0$ and $d_1$, and the resulting space and time exponents for the decoding phase. For simplicity, let us focus on the strongest and most natural attack, the \textit{interleaving attack}, where given $k$ ones and $c - k$ zeros in segment $i$, the collusion sets $y_i = 1$ with probability $k/c$ (i.e.\ $\theta_k = \frac{k}{c}$). Equivalently, for each segment the colluders random choose one of their members, and output his content. In that case $\expn_{p_i}(y_i) = p_i$ and we can further simplify the expressions for $\mu_0$ and $\mu_1$:
\begin{align}
    \mu_0 &= \expn_{\vc{p}, \vc{x}_j, \vc{y}}(s_j | H_0) = \ell \cdot \expn_p\left(\frac{p^2 + (1-p)^2 - 2 p (1-p)}{\sqrt{p(1-p)}}\right), \\
    \mu_1 &= \expn_{\vc{p}, \vc{x}_j, \vc{y}}(s_j | H_1) = \left(1 - \frac{1}{c}\right) \cdot \mu_0 + \frac{\ell}{c} \cdot \expn_p\left(\frac{1}{\sqrt{p(1-p)}}\right).
\end{align}
Using the optimal discrete distributions of~\cite{nuida09} for small $c$, optimized for the symmetric score function, and computing the resulting parameters, we obtain Table~\ref{tab:num}. Although the entire asymptotic trade-off spectrum is defined by Equation~\ref{eq:tradeoff} and $\alpha$, we explicitly instantiate these trade-offs in the last three columns, for the near-linear space regime ($\rho_s = 0$), the balanced regime ($\rho_q = \rho_s$), and the subpolynomial query time regime ($\rho_q = 0$). 

\subsection{Many colluders}
As one can see in the table, as $c$ increases the time and space exponents for the decoding phase quickly increase. For instance, for $c = 6$, we can obtain a query time complexity scaling as $n^{0.72}$, with space scaling as $n^{1.72}$, or if we insist on using only quasi-linear memory in $n$, the best query time complexity scales as $n^{0.97}$. 

For asymptotically large $c$, using the arcsine distribution with a cut-off $\delta > 0$ (optimally scaling as $\delta \propto c^{-4/3}$~\cite{laarhoven14dcc}), both the innocent and guilty scores scale such that, after normalization, we get $d_{0,1} \propto \delta^{1/4} \to 0$. The fact that $d_0 \approx d_1$ for large $c$ logically follows from the fact that colluders are able to blend in with the crowd better and better as $c$ increases, requiring large code length. Since $d_1/d_0 \to 1$, NNS techniques do not give any improvement in the limit of large $c$, and the main benefits are obtained when more memory is available to index the code words, and when $c$ is small. 

\newpage 
\section{Experiments}
\label{sec:exp}

To give an example of the potential speed-up in practice, we performed experiments for the interleaving attack with $c = 3$ colluders. In total we simulated $n = 10^5$ innocent users in each of $1000$ trials, where we used a code length of $\ell = 5000$. For the bias distribution we used the optimal $p$-values of~\cite{nuida09} of $p = 0.5 \pm 0.289$ with equal weights for both possibilities. 

\paragraph{NNS data structure.} For the NNS data structure and decoder, we implemented the asymptotically suboptimal but often more practical hyperplane locality-sensitive hashing method of Charikar~\cite{charikar02}. For the hyperplane LSH data structure, we chose the number of hash tables as $t = 100$, and we used a hash length of $k = 16$ (see~\cite{charikar02} for more details). For each of the $t$ hash tables, each of the $n$ data vectors is stored in one of the $2^k = 65536$ hash buckets. Given a query vector $\vc{q}$, for each of the $t$ hash tables we (1) compute $k$ inner products with random (sparse~\cite{achlioptas01}) unit vectors, with a total cost of $1600$ sparse dot products, and (2) do look-ups in these hash buckets for potential near neighbors (colluders), by computing their scores. 

\begin{figure}[t]
  \centering
  \includegraphics[width=\linewidth]{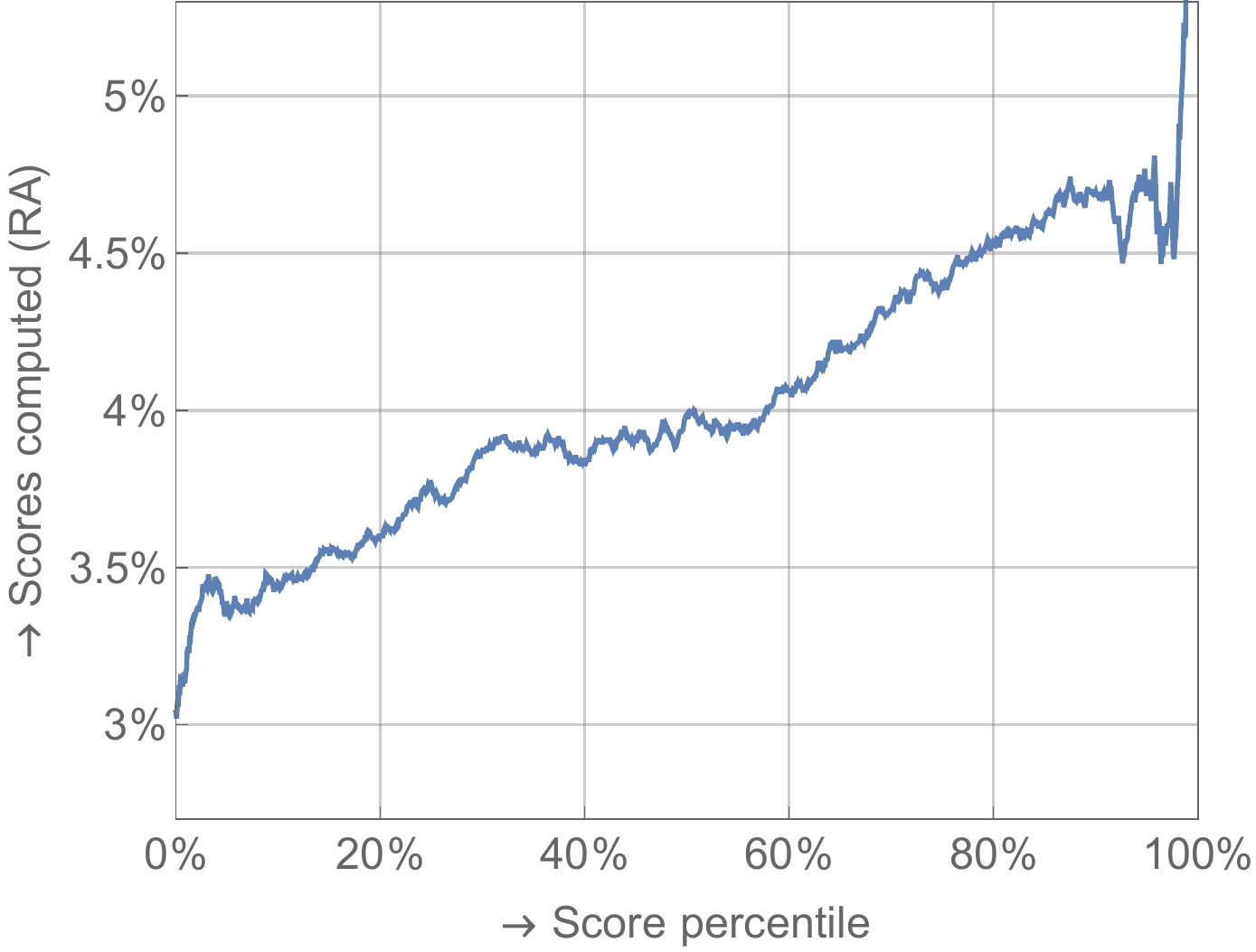}
  \caption{An illustration of (running averages of) how many user scores are computed, as a function of the user scores. \label{fig:exp}}
\end{figure}

\paragraph{Results.} Figure~\ref{fig:exp} illustrates (running averages of) how many user scores are commonly computed, i.e.\ how much work is done in the decoding stage, depending on how high the user scores are; the higher the score $s_j$, the larger the dot product $\langle \vc{v}_j, \vc{q} \rangle$, and the more likely it is we will find user $j$ (vector $\vc{v}_j$) colliding with $\vc{q}$ in one of the hash tables. From $1000$ simulations of the collusion process, approximately $4.2\%$ of all innocent users were considered as potential colluders (i.e.\ on average $4200$ of $10^5$ user scores were computed), and over $31\%$ of all colluders were found through collisions in the hash tables (i.e.\ on average approximately $1$ of the $3$ colluders was found). On average, the decoding consists of computing $1600$ dot products for the hash table look-ups, and $4200$ score computations of innocent users, for a total of around $5800$ dot products of length $\ell$. Compared to a naive linear search, which requires computing all $10^5$ user scores, the decoding is a factor $17$ faster. This comes at the cost of requiring $100$ hash tables, which each store pointers to all $n$ vectors in memory; since pointers are much smaller than the actual vectors, in practice the NNS data structure only required a factor $2$ more memory compared to no indexing.

\paragraph{Theory vs. Practice.} Theoretically, with $t = 100$ hash tables we are using of the order $t \cdot n = n^{1.40}$ memory, i.e.\ setting $\rho_s = 0.40$, (although in practice the memory only increases by a small amount). With $c = 3$, according to Table~\ref{tab:num} we have $\alpha \approx 1.22$ for the optimal asymptotic trade-offs, which according to~\eqref{eq:tradeoff} would thus result in $\rho_q \approx 0.58$. In reality the average query cost is computing $5800$ dot products, corresponding to a query exponent $\rho_q = \log(5800)/\log(10^5) \approx 0.75$. In practice, one may indeed notice that $\rho_q$ is slightly higher than the theoretical values suggest, but the memory increase is commonly much less than expected.

Note that although in most runs at least one colluder is successfully found in the hash tables (and has a large score), sometimes none of the colluders are found, and in some applications finding all colluders may be required. To get a higher success rate of finding colluders, one could for instance use multiprobing~\cite{andoni15cp} to still get a significantly lower decoding cost for finding all colluders compared to a linear search, without further increasing the memory.

\section{Discussion} 
\label{sec:disc}

Besides the main analysis on the costs of the decoding method, and the associated effect on the memory complexity, let us finally discuss a few more properties and aspects of the techniques outlined in this paper, which may affect how practical this method truly is.

\paragraph{Score function.} Although we only explicitly analyzed the application to the symmetric score function~\cite{skoric08}, the same techniques can be applied to any score-based scheme. For other score functions, the normalization factor $\vc{q}$ may depend on the attack strategy however, making an accurate instantiation of the NNS data structure harder unless the attack is known in advance.

\paragraph{Effects on encoding and embedding.} Even if the decoding method is more efficient, deploying this method in practice may not be cost-effective if the method for generating fingerprints and embedding these in the data needs to be modified. This is fortunately not a concern here, as the only thing that needs to be modified is how the owner of the content stores the code words $\vc{x}_j$ for decoding purposes: the \textit{exact same} encoding and embedding techniques can still be used.

\paragraph{Decoding accuracy.} One of the main reasons NNS techniques are fast, is that they allow for a small margin of error in the decoding procedure. In the application to fingerprinting, this means that the decoder may not always identify colluders from straightforward look-ups in the hash tables. This problem can be mitigated with multiprobing techniques~\cite{andoni15cp, panigrahy06}, or one could use NNS techniques without false negatives, such as~\cite{pagh16}.
    
\paragraph{Overhead of NNS techniques.} With NNS techniques, we reduce the asymptotic decoding time from $O(\ell n)$ to $O(\ell n^{\rho})$ with $\rho \leq 1$. Depending on how the hidden constants change, the effects may not immediately be visible for small $n$. Note however that the setting considered here, of solving NNS for data on the sphere, can be handled effectively with practical NNS techniques such as~\cite{charikar02, andoni15cp}, which have previously been proven to be much faster than linear searches on various benchmarks~\cite{aumueller17}, and our preliminary experiments confirm the improvement in practice.
    
\paragraph{Dynamic settings.} For streaming applications~\cite{fiat99, tassa05, laarhoven13tit, laarhoven15ihmmsec}, decisions on whether to accuse users or not need to be made in real-time as well. As the NNS techniques considered here commonly rely on static data, it is not directly obvious whether the same speed-ups can be obtained when the data arrives in a streaming fashion. Interested readers may consider~\cite{law05, liu03} for further reading on NNS techniques that may be relevant for streaming data.
    
\paragraph{Joint decoding.} While simple decoders with a decoding cost linear in $n$ might be considered reasonably efficient, in the joint decoding setting, the decoding cost of $O(\ell n^k)$ for $k \geq 2$ is a real problem. Similar techniques can be applied there, by slightly changing how the decoding problem is modeled as a near neighbor problem. In that case, the decoding time becomes $O(\ell n^{k \rho})$, and the improvement may be even more noticeable than for simple decoders.

\paragraph{Group testing.} As discussed in e.g.\ \cite{meerwald11b, laarhoven13allerton, skoric15b, laarhoven16phd, laarhoven14ihmmsec}, the group testing problem of detecting infected individuals among a large population using simultaneous testing~\cite{dorfman43}, is equivalent to the fingerprinting problem where the collusion strategy is fixed to the all-$1$ attack: whenever allowed by the marking assumption, the colluders output the symbol $1$. Similar techniques as described above can be applied there to reduce the decoding time complexity both for simple and joint group testing methods.

\begin{acks}
The author thanks Peter Roelse for discussions on the potential relevance of these techniques. The author is supported by an NWO Veni Grant under project number 016.Veni.192.005.
\end{acks}

\bibliographystyle{ACM-Reference-Format}
\bibliography{Database}

\end{document}